\documentclass{emulateapj} 
\usepackage{psfig}
\usepackage{apjfonts}
\usepackage{mathrsfs}
\usepackage{graphicx}

\begin{document}

\title{The origin of the spurious iron spread in the globular cluster NGC~3201
\footnote{Based on FLAMES observations collected 
under the Program 072.D-0777. }}

\author{A. Mucciarelli$^{1}$, E. Lapenna$^{1}$, D. Massari$^{1}$, 
F.R. Ferraro$^{1}$, B. Lanzoni$^{1}$}

\affil{$^{1}$Dipartimento di Fisica \& Astronomia, Alma Mater Studiorum, Universit\`a 
di Bologna, Viale Berti Pichat, 6/2 - 40127, Bologna, ITALY}

\begin{abstract}

NGC~3201 is a globular cluster suspected to have an intrinsic spread in  
the iron content.
We re-analysed a sample of 21 cluster stars observed with UVES-FLAMES at the Very Large 
Telescope and for which Simmerer et al. found a 0.4 dex wide [Fe/H] distribution with a 
metal-poor tail.
We confirmed that when spectroscopic gravities are adopted, the derived [Fe/H] distribution spans $\sim$0.4 dex.
On the other hand, when photometric gravities are used, the metallicity distribution 
from Fe~I lines remains large, while that derived from Fe~II lines is narrow and compatible 
with no iron spread.
We demonstrate that the metal-poor component claimed by Simmerer et al.
is composed by asymptotic giant branch stars that could be affected by non local thermodynamical
equilibrium effects driven by iron overionization. 
This leads to a decrease of the Fe~I abundance, while leaving the Fe~II abundance unaltered.
A similar finding has been already found in asymptotic giant branch 
stars of the globular clusters M5 and 47~Tucanae.
We conclude that NGC~3201 is a normal cluster, with no evidence of intrinsic iron spread.

\end{abstract}

\keywords{stars: abundances ---
techniques: spectroscopic ---
globular clusters: individual (NGC~3201)}   

\section{Introduction}
The striking homogeneity in terms of iron abundance is usually considered 
as the main chemical signature of globular clusters (GCs), indicating 
that these stellar systems were not able to retain the gas ejected by supernova (SN) explosions 
\citep[see e.g.][]{renzini86,willman12}.
For four decades, only one GC-like system, namely $\omega$ Centauri, was known 
to display an intrinsic dispersion in Fe 
\citep[see e.g.][]{freeman75,norris95,origlia03,johnson10,pancino11} 
and this evidence brought to classify the system as the remnant core of a tidally 
disrupted dwarf galaxy accreted by the Milky Way. 
In the last few years, deep and extensive spectroscopic and photometric investigations 
have revealed a more complex picture, with the discovery of other (massive) GCs harboring
distinct sub-populations with different iron abundance,  
as Terzan~5 \citep{ferraro09,origlia11,origlia13,massari14} and  M2 \citep{yong14}, or with large but 
uni-modal iron distributions, as M54 \citep{carretta10a} and M22 \citep{marino09}.
Further suggestions of intrinsic iron spread are only tentative at the moment, 
either because based on Ca~II lines \citep[NGC5824,][]{dacosta14} or because 
of still disputed results \citep[NGC1851,][]{carretta10b,villanova10}.

Among these candidate {\sl anomalous} GCs,
the case of NGC~3201 is controversial because different analyses 
provide conflicting results about its level of iron homogeneity. 
\citet{gonz98} first analysed a sample of CTIO high-resolution spectra of 18 cluster stars, 
finding large iron variations ($\Delta[Fe/H]\sim0.4$ dex). However, an evident trend 
between [Fe/H] and the effective temperature casts doubts about the reliability of their 
[Fe/H] distribution.
The re-analysis of six of their targets performed by \citet{covey03} does not 
provide additional clues.

Further analyses by \citet{carretta09} and \citet{munoz13}, based on high-resolution, 
high signal-to-noise ratio spectra (FLAMES@VLT and MIKE@Magellan, respectively), 
do not highlight similar spreads, ruling out large star-to-star variations.
On the other hand, \citet{simmerer13} analysed UVES@FLAMES and MIKE@Magellan high-resolution spectra 
of 24 giant stars of the cluster, revealing a metallicity distribution as large as 0.4 dex 
(not explainable within the uncertainties) and with an evident metal-poor tail (5 out 24 stars). 
This iron spread, qualitatively similar to that observed in M22 \citep{marino09}, would make NGC~3201 the 
least massive GC \citep[$\sim1.1\times10^5 M_{\odot}$;][]{mcl05} with evidence of SN ejecta retention.

Recently, \citet{lp14} discovered effects due to the departure from local 
thermodynamical equilibrium (NLTE) in a sample of asymptotic giant branch (AGB) stars 
in the globular cluster 47~Tucanae.
Such NLTE effects affect the abundances derived from Fe~I lines 
(bringing to an under-estimate of 0.1-0.2 dex in [Fe/H]), but leave 
the abundances from Fe~II lines unaltered. On the other hand, red giant branch 
(RGB) stars do not exhibit similar effects and the abundances from Fe~I and Fe~II nicely match each other.

Based on this finding, it is well reasonable to ask whether the star-to-star scatter measured 
in the Fe content of NGC~3201 is genuine or it is due to such a spurious effect, because of the inclusion 
in the sample of some AGB stars. Note that the NLTE effects can be easily unveiled by comparing 
the abundances derived from Fe~I and Fe~II separately.

In this paper we re-analyse the spectra of the stars discussed in \citet{simmerer13} 
in light of the finding by \citet{lp14}. We limit the analysis to the UVES-FLAMES sample, 
including also the 5 stars populating the metal-poor tail of the \citet{simmerer13} 
distribution.

\section{Observations}
High-resolution spectra taken with UVES-FLAMES@VLT \citep{pasquini} for 21 giant stars members of NGC~3201
have been retrieved from the ESO archive. The spectra have been acquired with the UVES grating 580 
Red Arm CD\#3, that provides a high spectral resolution (R$\sim$45000) and a large spectral coverage 
($\sim$4800-6800 \AA). The spectra have been reduced using the dedicated ESO
pipeline\footnote{http://www.eso.org/sci/software/pipelines/}, performing bias subtraction, flat-fielding, 
wavelength calibration, spectral extraction and order merging. In each exposure one fiber is dedicated 
to sample the sky background and used to subtract this contribution from each individual spectrum.

Spectroscopic targets have been identified in our photometric catalog, obtained by combining 
high resolution images acquired with the HST-ACS camera and wide-field images 
acquired with the ESO-WFI imager. Both the photometric datasets have been obtained through the V and 
I filters. A total of 13 targets lie in the innermost cluster region, covered by ACS, while 8 stars are 
in the external region, covered by WFI.
The membership of all the targets is confirmed by their very high radial velocity 
($<RV_{helio}>=+494.6\pm0.8$ km s$^{-1}$, $\sigma=3.6$ km s$^{-1}$) that allows to easily distinguish 
the cluster members from the surrounding field stars.

\begin{figure*}
\plottwo{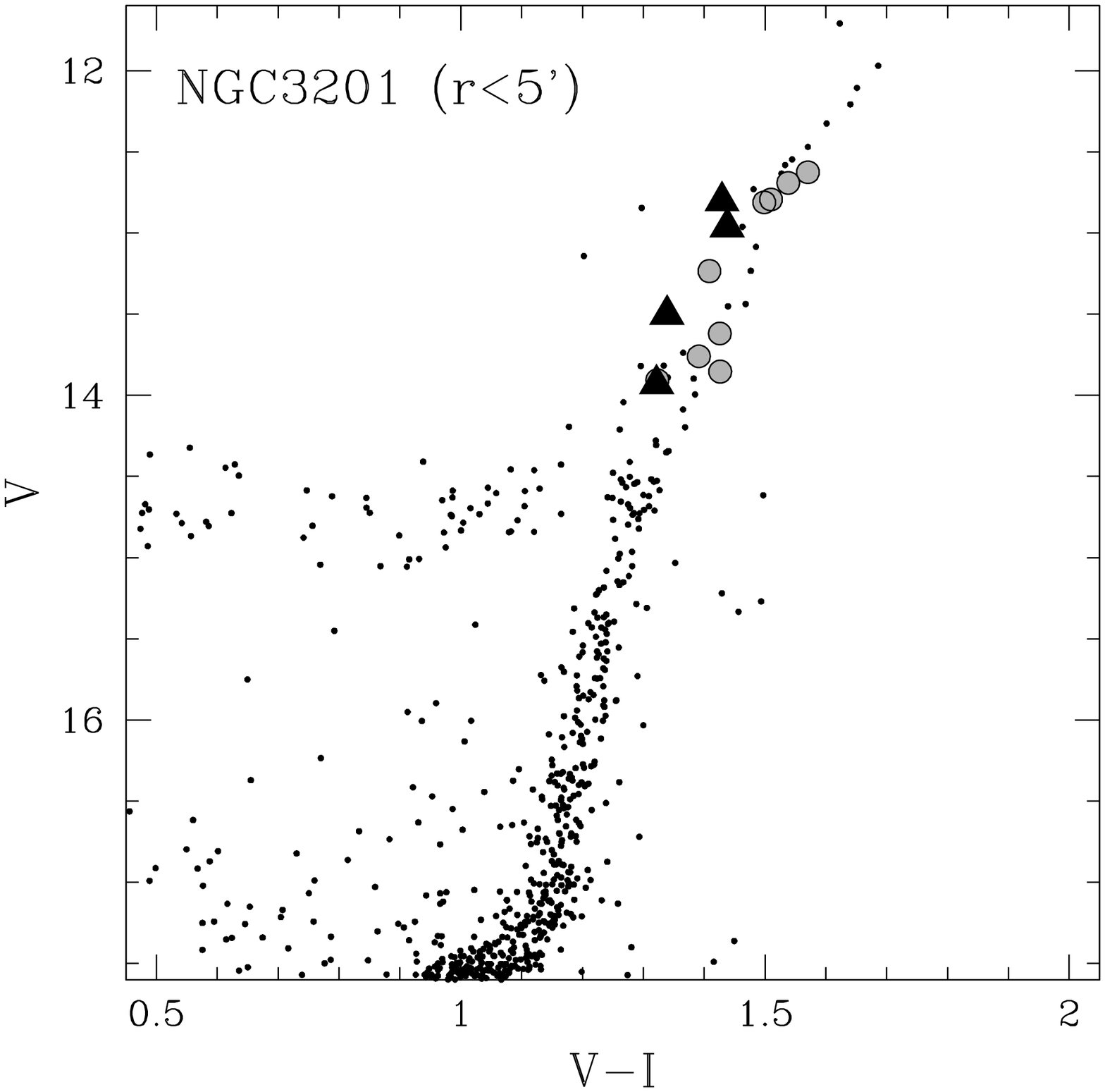}{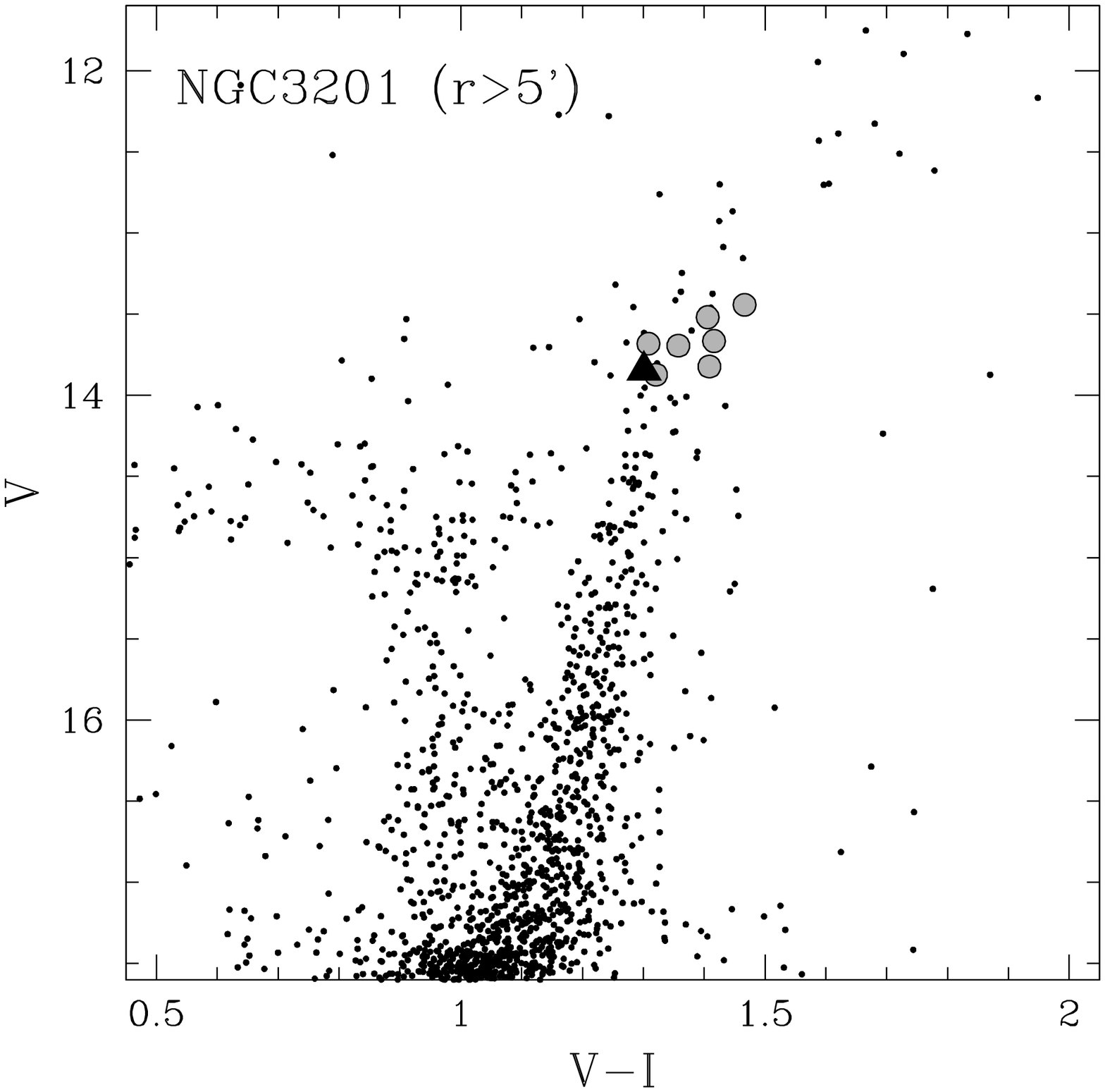}
\caption{CMDs for central and external regions of NGC~3201 
(left and right panels, respectively), 
corrected for differential reddening. Large circles are the spectroscopic 
targets flagged as {\sl metal-rich} ([Fe/H]$>$--1.58 dex) in \citet{simmerer13}, while 
large triangles those identified as {\sl metal-poor} ([Fe/H]$\le$--1.58 dex).}
\label{cmd}
\end{figure*}

The position of the targets in the color-magnitude diagrams (CMDs) is shown in Fig.~\ref{cmd}.
These CMDs have been corrected for differential reddening using the method described 
in \citet{massari12} 
and adopting the extinction law by \citet{cardelli89}. 
In order to calculate guess values for the atmospheric parameters of the target stars, 
we fitted the CMDs with 
an appropriate theoretical isochrone from the BaSTI dataset 
\citep{pietrinferni06}, computed with an age of 11 Gyr \citep{marin}, 
Z=~0.001 and $\alpha$-enhanced chemical mixture, finding a color excess E(B-V)=~0.31 mag and 
a true distance modulus $(m-M)_0$=13.35 mag.

Table 1 lists the main information about the targets, by adopting the same 
identification numbers used by \citet{simmerer13} who adopted the original names by \citet{cote94}.

\begin{deluxetable*}{lcccccccccc}
\tablecolumns{11} 
\tablewidth{0pc}  
\tablecaption{}
\tablehead{ 
\colhead{Star} &   RA & Dec &  V  &  I & $RV_{helio}$    & $T_{\rm eff}$ &  {\sl log~g} & $v_{\rm turb}$  &  [Fe~I/H] & [Fe~II/H]\\
  &   (J2000)  &  (J2000) &  &   & (km/s)  & (K)  &   & (km/s) & (dex) & (dex)}
\startdata 
\hline 
 63 &	154.3084680  &   -46.4125790  &  13.76  &  12.42  & 495.9$\pm$0.6        &  4730 &  1.54	&   1.45     &  -1.27$\pm$0.03   & -1.35$\pm$0.05  \\	 
 89 &	154.3346190  &   -46.3860090  &  13.87  &  12.56  & 498.8$\pm$0.5        &  4855 &  1.63	&   1.60     &  -1.57$\pm$0.02   & -1.35$\pm$0.03  \\	   
 91 &	154.3359600  &   -46.3355290  &  13.81  &  12.41  & 490.4$\pm$0.6        &  4705 &  1.56	&   1.50     &  -1.36$\pm$0.05   & -1.35$\pm$0.05  \\	 
105 &	154.3439854  &   -46.4201000  &  13.94  &  12.59  & 493.9$\pm$0.9        &  4760 &  1.61	&   1.45     &  -1.31$\pm$0.03   & -1.29$\pm$0.05  \\	
124 &	154.3601786  &   -46.4140432  &  12.85  &  11.32  & 498.5$\pm$0.2        &  4375 &  0.96	&   1.55     &  -1.37$\pm$0.02   & -1.40$\pm$0.05  \\	
129 &	154.3623900  &   -46.4312480  &  13.54  &  12.13  & 491.8$\pm$0.4        &  4580 &  1.36	&   1.50     &  -1.48$\pm$0.02   & -1.43$\pm$0.04  \\	
181 &	154.3876490  &   -46.4126379  &  13.89  &  12.58  & 497.5$\pm$0.4        &  4920 &  1.65	&   1.65     &  -1.62$\pm$0.03   & -1.37$\pm$0.04  \\  
200 &	154.3949420  &   -46.3972062  &  12.79  &  11.30  & 495.1$\pm$0.4        &  4515 &  0.99	&   1.85     &  -1.50$\pm$0.02   & -1.40$\pm$0.04  \\  
222 &	154.4031199  &   -46.4257714  &  12.69  &  11.15  & 491.0$\pm$0.5        &  4355 &  0.88	&   1.65     &  -1.47$\pm$0.02   & -1.45$\pm$0.05  \\  
231 &	154.4067690  &   -46.4012923  &  13.52  &  12.18  & 491.2$\pm$0.5        &  4785 &  1.45	&   1.60     &  -1.54$\pm$0.02   & -1.39$\pm$0.03  \\	  
240 &	154.4091582  &   -46.4277707  &  13.91  &  12.59  & 495.2$\pm$0.6        &  4855 &  1.64	&   1.55     &  -1.61$\pm$0.03   & -1.38$\pm$0.04  \\	 
244 &	154.4097946  &   -46.4024205  &  13.77  &  12.38  & 492.2$\pm$0.3        &  4690 &  1.51	&   1.50     &  -1.44$\pm$0.02   & -1.44$\pm$0.05  \\	   
249 &	154.4104867  &   -46.4270988  &  12.99  &  11.54  & 498.1$\pm$0.3        &  4545 &  1.11	&   1.60     &  -1.56$\pm$0.02   & -1.42$\pm$0.04  \\	 
277 &	154.4198397  &   -46.4124756  &  13.60  &  12.18  & 497.6$\pm$0.3        &  4620 &  1.39	&   1.50     &  -1.40$\pm$0.02   & -1.42$\pm$0.05  \\	
279 &	154.4208433  &   -46.3954964  &  13.34  &  11.89  & 499.0$\pm$0.3        &  4555 &  1.26	&   1.50     &  -1.43$\pm$0.03   & -1.43$\pm$0.04  \\	
303 &	154.4288640  &   -46.3800960  &  13.58  &  12.20  & 491.7$\pm$0.4        &  4630 &  1.39	&   1.50     &  -1.45$\pm$0.02   & -1.45$\pm$0.04  \\	
308 &	154.4304349  &   -46.4108231  &  13.76  &  12.37  & 487.2$\pm$0.4        &  4620 &  1.45	&   1.45     &  -1.50$\pm$0.02   & -1.45$\pm$0.04  \\  
312 &	154.4342122  &   -46.4240651  &  12.71  &  11.10  & 494.9$\pm$0.3        &  4330 &  0.85	&   1.65     &  -1.44$\pm$0.03   & -1.44$\pm$0.06  \\  
332 &	154.4478331  &   -46.3982898  &  12.96  &  11.47  & 499.7$\pm$0.5        &  4500 &  1.08	&   1.65     &  -1.54$\pm$0.02   & -1.38$\pm$0.04  \\  
344 &	154.4544130  &   -46.4162830  &  13.82  &  12.41  & 490.7$\pm$0.7        &  4655 &  1.49	&   1.40     &  -1.31$\pm$0.03   & -1.29$\pm$0.05  \\  
374 &	154.5004800  &   -46.5194470  &  13.64  &  12.09  & 498.7$\pm$0.6        &  4595 &  1.38	&   1.50     &  -1.41$\pm$0.03   & -1.36$\pm$0.05  \\	  
\hline
    &	             &                &      &  &             &       &         &            &  $<$[Fe~I/H]$>$        &    $<$[Fe~II/H]$>$     \\	 
    &	             &                &      &  &              &       &         &            &   --1.46$\pm$0.02      &   --1.40$\pm$0.01  
\enddata 
\tablecomments{$~~~~~$Main information of the target stars. 
Identification numbers are the same adopted by \citet{simmerer13}.
[Fe I/H] and [Fe II/H] have been obtained adopting photometric gravities.}
\end{deluxetable*}

\section{Analysis}

Iron abundances have been derived with the package 
GALA\footnote{http://www.cosmic-lab.eu/gala/gala.php}\citep{mucciarelli13}
by matching the measured and theoretical equivalent widths (EWs).
Model atmospheres have been calculated with the code 
ATLAS9\footnote{http://wwwuser.oats.inaf.it/castelli/sources/atlas9codes.html}.
We selected Fe~I and Fe~II lines predicted to be unblended at the UVES resolution and at the 
typical atmospheric parameters and metallicity of the observed stars, through the careful inspection 
of synthetic spectra calculated with the SYNTHE package \citep{sbordone05}. Atomic data 
of the transitions of interest are from the last release of the Kurucz/Castelli linelist
\footnote{http://wwwuser.oats.inaf.it/castelli/linelists.html}. The final iron abundances are based 
on $\sim$130-150 Fe~I and $\sim$15-20 Fe~II lines.
EWs have been measured with DAOSPEC \citep{stetson08}, run iteratively  
by means of the package 4DAO\footnote{http://www.cosmic-lab.eu/4dao/4dao.php}\citep{mucciarelli13b}.
EW, oscillator strength and excitation potential for all the measured transitions are listed 
in Table 2 (available in its entirety in the online version).

\begin{deluxetable}{cccccc}
\tablecolumns{6} 
\tablewidth{0pc}  
\tablecaption{Star identification number, wavelength, oscillator strength, excitation potential and 
measured EWs for all the used transitions.}
\tablehead{ 
\colhead{Star} &   $\lambda$ & Ion &  log(gf)  & E.P. & EW   \\
  &   (\AA) &  &  & (eV)  & (m\AA)}
\startdata 
\hline 
 63  &   4791.246   &	FeI       &  -2.435  &  3.270  &  26.40     \\    
 63  &   4834.507   &	FeI  	  &  -3.330  &  2.420  &  27.90     \\      
 63  &   4842.788   &	FeI  	  &  -1.530  &  4.100  &  16.30     \\    
 63  &   4892.859   &	FeI  	  &  -1.290  &  4.220  &  29.50     \\   
 63  &   4911.779   &	FeI  	  &  -1.760  &  3.930  &  19.50     \\   
 63  &   4917.230   &	FeI  	  &  -1.160  &  4.190  &  35.90     \\   
 63  &   4918.013   &	FeI  	  &  -1.340  &  4.230  &  28.30     \\  
 63  &	 4950.106   &	FeI  	  &  -1.670  &  3.420  &  54.10    \\  
 63  &	 4962.572   &	FeI  	  &  -1.182  &  4.180  &  32.30    \\  
 63  &	 4969.917   &	FeI  	  &  -0.710  &  4.220  &  46.00    \\	  
 63  &	 4985.253   &	FeI  	  &  -0.560  &  3.930  &  75.30    \\	 
 63  &	 5002.793   &	FeI  	  &  -1.530  &  3.400  &  67.20    \\	   
 63  &	 5014.942   &	FeI  	  &  -0.303  &  3.940  &  86.70    \\	 
 63  &	 5022.236   &	FeI  	  &  -0.560  &  3.980  &  76.40    \\	
 63  &   5028.126   &	FeI  	  &  -1.123  &  3.570  &  68.30     \\  
\hline
\enddata 
\tablecomments{$~~~~~$This table is available in its entirety in a machine-readable form 
in the online journal. A portion is shown here for guidance regarding its form and content.}
\end{deluxetable}

\subsection{Analysis with spectroscopic gravities} 
First, we performed a fully spectroscopic analysis, as done by 
\citet{simmerer13}, in order to verify whether we obtain the same evidence of 
a metallicity dispersion. In this analysis the atmospheric parameters have been constrained 
as follows:
{\sl (a)}~for the effective temperatures ($T_{\rm eff}$) we requested that no trend exists between abundances 
and excitation potential, 
{\sl (b)}~for the surface gravities ({\sl log~g}) we imposed that the same abundance is obtained 
(within the uncertainties) from Fe~I and Fe~II lines, 
{\sl (c)}~for the microturbulent velocity ($v_{\rm turb}$) we requested that no trend exists between abundances 
from Fe~I lines and the reduced line strength. The derived values of $v_{\rm turb}$ are based on $\sim$130-150 Fe~I lines 
distributed over a large interval of reduced EWs, with log(EW/$\lambda$) ranging between --5.6 and --4.7.

We derived an average [Fe~I/H]=--1.46$\pm$0.02 dex ($\sigma$=~0.10 dex), with a 
distribution ranging from --1.62 dex to --1.27 dex. 
The [Fe~I/H] and [Fe~II/H] abundance distributions are shown in Fig.~\ref{histo} 
as generalized histograms.
This result well matches that obtained by \citet{simmerer13} that find an average abundance 
[Fe/H]=--1.48$\pm$0.02 dex ($\sigma$=~0.11 dex)\footnote{Note that \citet{simmerer13} adopted as solar reference value 
7.56 obtained from their own solar analysis, while we used 7.50 by \citet{gs98}. 
Througout the paper we refer to the abundances by \citet{simmerer13} corrected 
for the different solar zero-point.} with a comparable iron range ($\Delta[Fe/H]\sim$0.4 dex). 
This spectroscopic analysis fully confirms the claim by \citet{simmerer13}: when 
analysed with atmospheric parameters derived following the constraints listed above, 
the stars of NGC~3201 reveal a clear star-to-star scatter in the iron content.
The 5 stars labelled as {\sl metal-poor} 
by \citet{simmerer13}, with [Fe/H]$<$--1.58 dex, are the most metal-poor also in our 
metallicity distribution.

\subsection{Analysis with photometric gravities} 
As pointed out by \citet{lp14}, possible NLTE effects in AGB stars 
can be easily detected by assuming photometric values for {\sl log~g} and measuring
Fe~I and Fe~II independently\footnote{Recently, \citet{johnson14} analysed a sample 
of 35 AGB stars in 47 Tucanae, finding no clear evidence of NLTE effects.
Even if a detailed comparison between the two analyses is not the scope of this paper, 
we highlight some main differences between the two works: the spectral resolution 
\citep[48000 in \citet{lp14} and 22000 in][]{johnson14}, the number of Fe~II lines 
\citep[13 in \citet{lp14} and 4 in][, on average]{johnson14}
and the adopted linelists for AGB and RGB stars 
(\citet{lp14} adopted the same linelist for both the groups of stars, at variance 
with \citet{johnson14} that used a linelist consistent, but not exactly the same, 
with that adopted by \citet{cordero14} where the reference RGB stars are discussed).}.
In order to see whether this effect is present in the NGC~3201 data, we adopted the 
following procedure.
$T_{\rm eff}$ has been derived spectroscopically, by imposing the 
excitation equilibrium as described above. Thanks to the high quality 
of the spectra (with S/N ratio per pixel higher than 100) and the 
large number of Fe~I lines distributed over a large range of excitation potentials, 
very accurate spectroscopic $T_{\rm eff}$ can be estimated, with internal uncertainties 
of about 20-30 K.
As a guess value we adopted $T_{\rm eff}$ calculated from the 
$(V-I)_0 - T_{\rm eff}$ calibration by \citet{alonso99} and 
assuming a color excess E(B-V)=0.31 mag.
Gravities have been derived from the Stefan-Boltzmann equation,
assuming E(B-V)=0.31 mag, $(m-M)_0$=13.35 mag, 
bolometric corrections from \citet{alonso99} and a mass of 0.82 $M_{\odot}$ 
(according to the best-fit isochrone, the latter is 
a suitable value for RGB stars brighter than the RGB Bump magnitude level).
Because $T_{\rm eff}$ is derived spectroscopically, 
the gravity is recomputed through the Stefan-Boltzmann equation in each iteration 
according to the new value of $T_{\rm eff}$.
The mass value of 0.82 $M_{\odot}$ is appropriate for RGB stars 
but probably too high for AGB stars, because of mass loss phenomena during 
the RGB phase \citep{rood73,origlia02,origlia07,origlia14}. 
In fact, \citet{gratton10} provide the masses for a sample of HB stars in NGC~3201, 
finding values between 0.62 and 0.71 $M_{\odot}$. We initially 
analysed all the targets assuming the mass of a RGB star. Then, the AGB candidates, 
selected according to their position in $T_{\rm eff}$--log~g plane (as discussed in Section 4)
have been re-analysed by assuming the median value 
(0.68 $M_{\odot}$) of the HB stars estimated by \citet{gratton10}.

This method allows us to take advantage of the high-quality of the spectra, 
deriving accurate $T_{\rm eff}$ thanks to the large number of transitions 
spanning a large range of excitation potentials. On the other hand, this approach 
does not require a fully spectroscopic determination of {\sl log~g}
that is instead calculated using both photometric information and 
spectroscopic $T_{\rm eff}$, avoiding any possible bias 
related to NLTE effects.

\begin{figure*}
\plottwo{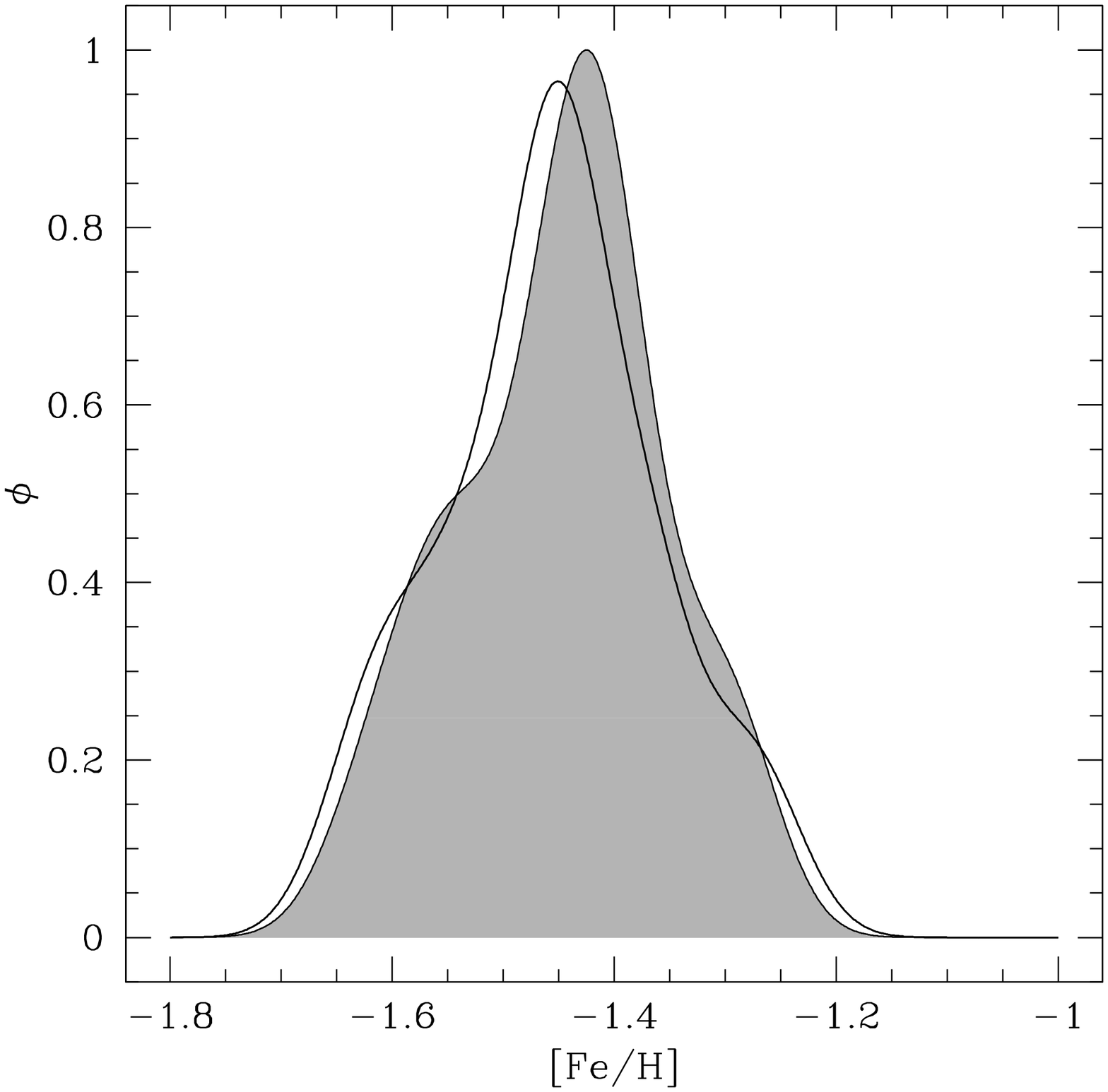}{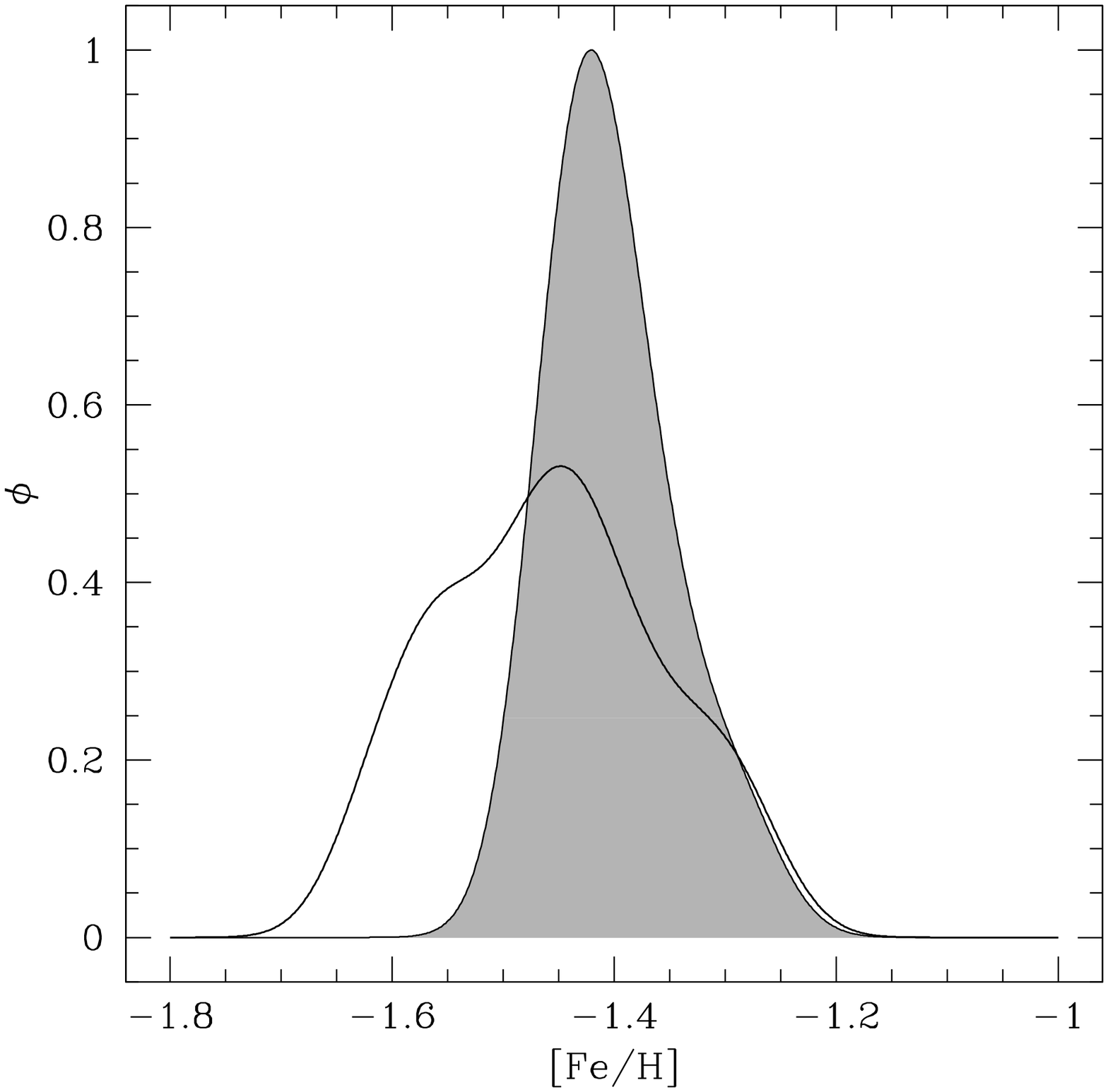}
\caption{Generalized histograms for [Fe~I/H] (empty histogram) and [Fe~II/H] (grey histogram) 
obtained from the analysis performed with spectroscopic gravities (left panel) and with
photometric gravities (right panel).}
\label{histo}
\end{figure*}

The [Fe~I/H] and [Fe~II/H] abundances obtained with this method are listed in Table 1.
The right panel of Fig.~\ref{histo} shows the [Fe~I/H] and [Fe~II/H] distributions represented as generalized 
histograms obtained from this analysis. The two distributions turn out to be quite 
different. The iron distribution obtained from Fe~I lines resembles that obtained with the spectroscopic 
parameters (left panel of Fig.~\ref{histo}), with an average value of  [Fe I/H]=--1.46$\pm$0.02 dex
($\sigma$=~0.10 dex),
while the distribution obtained from Fe~II lines has a narrow gaussian-shape 
([Fe II/H]$>$=--1.40$\pm$0.01 dex, $\sigma$=~0.05 dex) pointing to a quite 
homogeneous iron content.

\subsection{Uncertanties}
 
Internal uncertainties in the derived Fe abundances have been calculated by adding in quadrature 
two sources of uncertainties:\\
(1)~those arising from the EW measurement. For each target, we estimated this term as the line-to-line dispersion normalized 
to the root mean square of the number of lines. Because of the high quality of the used spectra, the 
line-to-line scatters are smaller than 0.1 dex, leading to internal uncertainties of about 0.005-0.008 dex 
for [Fe I/H] and of about 0.010-0.025 dex for [Fe II/H].\\ 
(2)~those arising from the atmospheric parameters. To estimate this term, we follow the approach described by 
\citet{cayrel04} to take into account the covariance terms due to the correlations among the atmospheric parameters.
For each target, the temperature has been varied by $\pm1\sigma_{T_{\rm eff}}$, the gravity has been 
re-calculated through the Stefan-Boltzmann equation 
adopting the new values of $T_{\rm eff}$ and the microturbulent velocity re-optmized spectroscopically.\\
Table 1 lists the total uncertainty including both the terms (1) and (2). 
Also, Table 3 shows for two representative targets (one RGB and one AGB star) the 
abundance uncertainty obtained following the prescriptions by \citet{cayrel04} (second column) and 
those obtained with the usual method of independently varying each parameter 
(an approach that obviously does not take into account the correlation among the parameters and 
can over-estimate the total uncertainty).

\begin{deluxetable}{ccccc}
\tablecolumns{5} 
\tablewidth{0pc}  
\tablecaption{Abundance uncertanties due to the atmospheric parameters 
for the stars \#63 and \#89.}
\tablehead{ 
\colhead{Ion} &   Parameters    & $\delta T_{\rm eff}$ &  $\delta$log~g  & $\delta v_{\rm turb}$    \\
              &   Uncertainty &   $\pm$50 K     &    $\pm$0.1     &  $\pm$0.1 km/s \\
	      &     (dex)       &     (dex)       &    (dex)        &   (dex)  }
\startdata 
\hline 
  &  &  \#63 (RGB) &  & \\
\hline 
 Fe~I   & $\pm$0.03 & $\pm$0.05 & $\pm$0.00 & $\mp$0.03 \\
 Fe~II  & $\pm$0.04 & $\mp$0.02 & $\pm$0.04 & $\pm$0.03 \\
\hline
  &  &  \#89 (AGB) &  & \\
\hline 
 Fe~I   & $\pm$0.02 & $\pm$0.04 & $\pm$0.00 & $\mp$0.03 \\
 Fe~II  & $\pm$0.03 & $\mp$0.02 & $\pm$0.04 & $\pm$0.02 
\enddata 
\tablecomments{$~~~~~$ The second column is the total uncertainty 
calculated according to \citet{cayrel04}. The other columns list the 
abundance variations related to the variation of only one parameter.}
\end{deluxetable}

\section{Discussion}

In this paper we present a new analysis of the UVES-FLAMES spectra of 21 member stars of NGC~3201 
already discussed in \citet{simmerer13}. 
The Fe abundances have been calculated both using spectroscopic gravities 
(obtained by imposing the ionization balance between Fe~I and Fe~II abundances) and 
photometric ones (obtained through the Stefan-Boltzmann equation). 
The two methods provide different results concerning [Fe~I/H] and [Fe~II/H]. 
In particular, the use of spectroscopic {\sl log~g} provides a wide [Fe/H] distribution 
(al large as $\sim$0.4 dex), in agreement with the finding of \citet{simmerer13} 
who adopted the same method. On the other hand, when photometric gravities are used,
the [Fe~I/H] distribution remains quite large, while that of [Fe~II/H] is narrow. 
We compute the intrinsic spread of the two Fe distributions adopting the Maximum Likelihood 
algorithm described in \citet{mucciarelli12}. Concerning [Fe~I/H] we derive $\sigma_{int}=0.09\pm0.01$ dex, 
while for [Fe~II/H] $\sigma_{int}=0.00\pm0.02$ dex. Hence, the [Fe~II/H] distribution is compatible with 
no iron spread.

\begin{figure*}
\plottwo{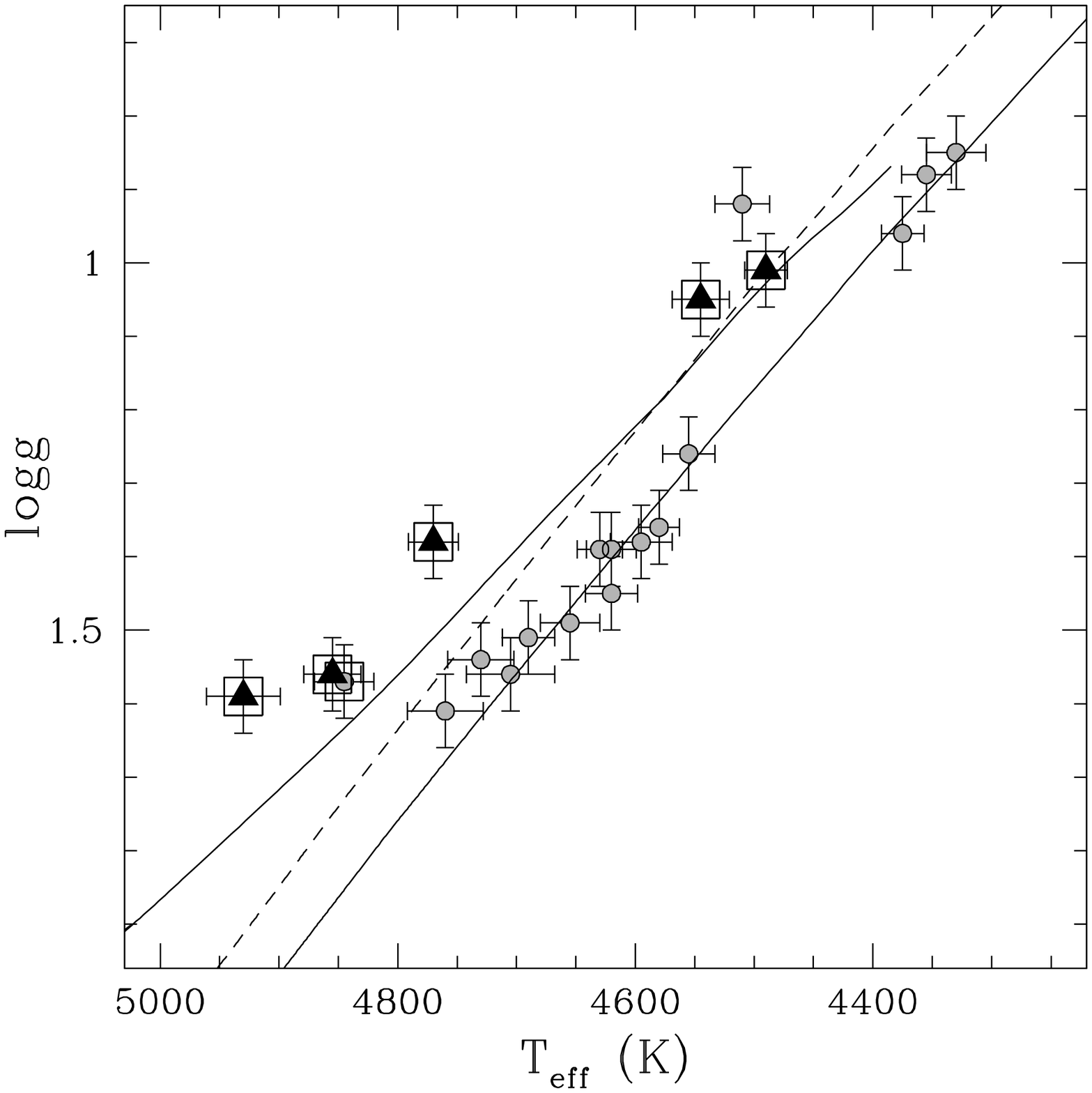}{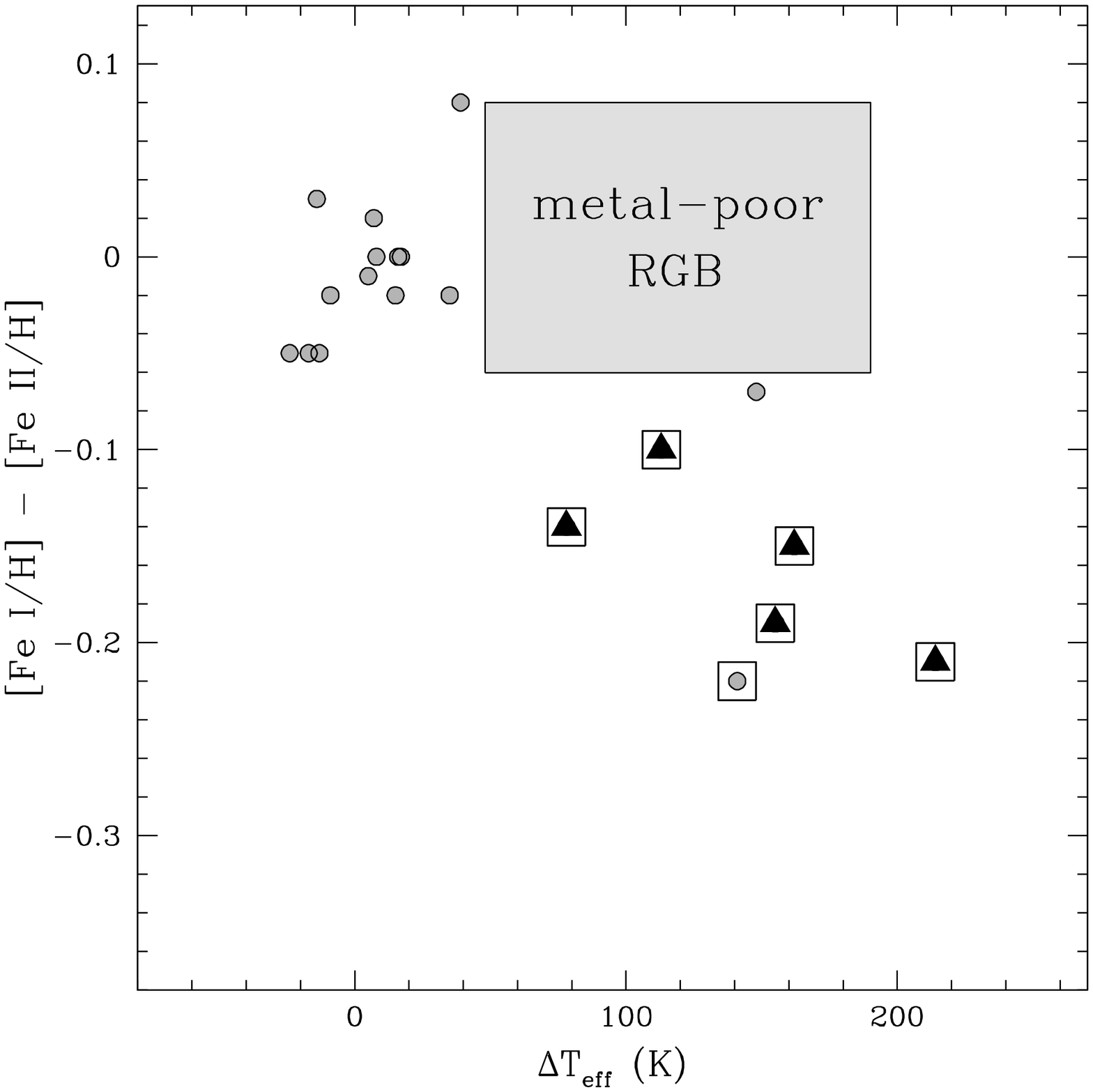}
\caption{{\sl Left panel}: position of the targets in the $T_{\rm eff}$--{\sl log~g} plane. 
A BaSTI isochrone with age of 11 Gyr, Z=0.001 and $\alpha$-enhancement chemical 
mixture (solid line) is superimposed for sake of comparison.
The dashed line indicates the position of the RGB for a BaSTI isochrone with age of 11 Gyr 
an Z=0.0006.
Grey triangles are the metal-poor stars of \citet{simmerer13}. 
Empty grey squares are the stars in our analysis with [Fe~I/H] - [Fe~II/H]$<$--0.1 dex. 
{\sl Right panel}: Behaviour of [Fe~I/H] - [Fe~II/H] as a function of the 
difference between the spectroscopic temperature of each target and the photometric 
value estimate from the isochrone with Z=0.001 shown in the left panel. 
Same symbols as in the left panel. The grey light area indicates the expected mean locus 
for metal-poor ([Fe/H]$\sim$--1.6 dex) RGB stars.}
\label{iso}
\end{figure*}

\citet{simmerer13} highlight that the 5 most metal-poor stars of their sample are bluer
than the other stars, as expected in cases of a lower metallicity. 
The left panel of Fig.~\ref{iso} shows the position of the targets in the $T_{\rm eff}$--{\sl log~g} plane, 
with superimposed, as reference, two isochrones with the same age but different 
metallicity: Z=~0.001 (solid line) and Z=~0.0006 (dashed line). 
The RGB of the isochrone with Z=~0.0006 overlaps the position of the AGB 
of the Z=~0.001 isochrone. 
Seven targets (including the 5 candidate metal-poor stars) are located 
in a position compatible with both the scenarios: metal-poorer RGB or AGB at the 
cluster metallicity. 
These seven targets have average abundances of [Fe I/H]=--1.57$\pm$0.01 dex and 
[Fe II/H]=--1.41$\pm$0.01 dex, while the RGB stars have [Fe I/H]=--1.42$\pm$0.02 dex and 
[Fe II/H]=--1.40$\pm$0.01 dex.

However, if these stars were metal-poor RGB stars,
the iron abundance derived from Fe~I lines should be in agreement 
with that obtained from Fe~II, since NLTE effects are not observed 
in RGB stars of comparable luminosity and metallicity. 
The right panel of Fig.~\ref{iso} shows the behaviour of ([Fe I/H] - [Fe II/H]) as a function 
of the difference between the spectroscopic values of $T_{\rm eff}$ and those obtained from 
the projection along the RGB of the best-fit isochrone.
The stars located along the RGB ($\Delta T_{\rm eff}\sim$0) have similar [Fe I/H] and [Fe II/H] 
abundances, compatible with no NLTE effects. The stars hotter than the 
reference RGB have differences between [Fe I/H] and [Fe II/H] 
ranging from --0.07 to --0.22 dex, with a mean value of --0.15 dex.
The grey region marks the expected position for metal-poorer RGB stars: they should be 
hotter than the reference RGB, but with [Fe I/H] - [Fe II/H]$\sim$0, as commonly measured 
in the RGB stars.
This reveals the true nature of these stars: they are genuine AGB stars, with 
the same metallicity of the cluster (as measured from their Fe~II lines) 
but affected by NLTE effects leading to a systematic decrease of [Fe~I/H]. 
This is the same effect observed by \citet{ivans01} and \citet{lp14} in the AGB stars 
of M5 and 47~Tucanae, respectively.

\begin{figure*}
\includegraphics[angle=-90,scale=0.65]{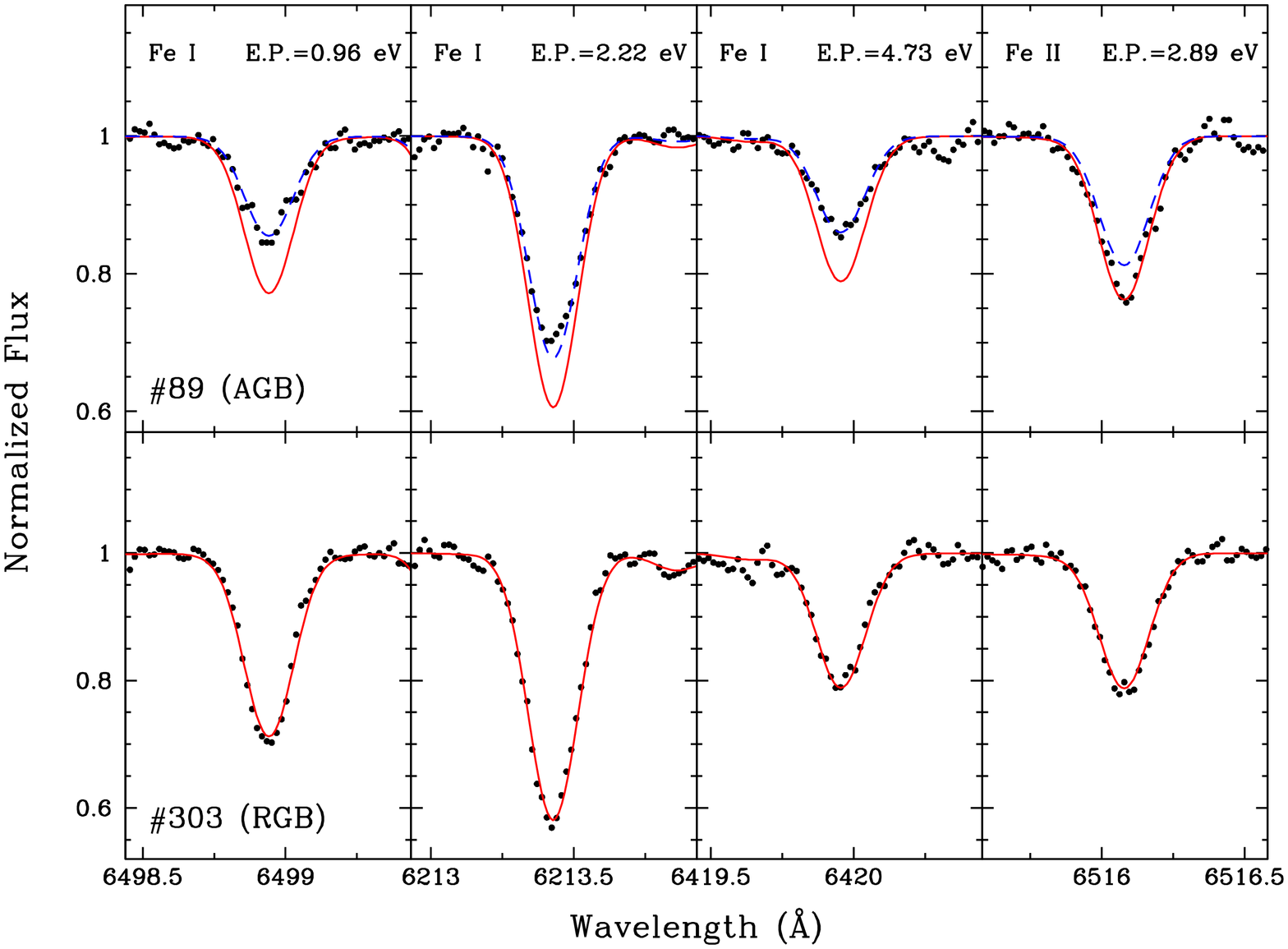}
\caption{Spectral regions around three Fe I lines with different excitation potential and 
one Fe II line, for the AGB star \#89 (upper panels) and the RGB star \#303 (lower panels). 
Synthetic spectra calculated with the corresponding atmospheric parameters (see Table 1) 
and adopting the average iron abundance derived from Fe II lines are superimposed as red curves. 
The blue dashed curve shown in the upper panels 
is the synthetic spectrum calculated with the iron abundance derived from Fe I lines.}
\label{spec}
\end{figure*}

A direct inspection of the spectra reveals the different behaviour of Fe I and Fe II lines 
in AGB and RGB stars. Fig.~\ref{spec} shows three Fe I lines (chosen with different excitation 
potential) and one Fe II line in the spectra of the AGB star \#89 (upper panels) and of the 
RGB star \#303 (lower panels). Synthetic spectra calculated with the appropriate atmospheric parameters 
and the metallicity derived from Fe~II lines (red lines).
In the upper panels we also show the synthetic spectrum computed with 
the average [Fe I/H] (blue dashed line).
Clearly, the synthetic spectrum assuming the [Fe~II/H] abundance well reproduces all the 
observed lines in the case of the RGB star, while it fails to fit the Fe~I lines observed in the 
AGB star, regardless of the excitation potential (pointing out that 
this effect cannot be attributed to inadequacies in the adopted $T_{\rm eff}$). On the other hand, the abundance derived 
from Fe~I lines is too low to well reproduce the depth of the AGB Fe~II line plotted in 
Fig.~\ref{spec}. This clearly demonstrates a different behaviour of iron lines in AGB 
and RGB stars.

It is worth noting that this behaviour is somewhat puzzling, because theoretical models do not 
predict significant differences in the NLTE corrections for stars in the parameter space 
covered by our targets. For instance, the grid of NLTE corrections computed by \citet{bergemann12} and 
\citet{lind12} predicts that the Fe~I lines in AGB and RGB stars should be affected 
in a very similar way at the metallicity of NGC~3201. However, some additional effects/mechanisms could play a role 
in the AGB photospheres, leading to the departure from the LTE condition, which are not yet 
accounted for in the available theoretical calculations.

We checked whether the mass assumed for AGB stars could change our conclusions.
The mass distribution of HB stars in NGC~3201 provided by \citet{gratton10} ranges from 0.62 to 
0.71 $M_{\odot}$, with a median value of 0.68 $M_{\odot}$. 
The minimum and maximum mass values correspond to a difference in  
log~g of 0.06, leading to a variation in [Fe II/H] of only 0.02-0.025 dex 
(and a variation in [Fe I/H] of 0.002-0.005 dex).
On the other hand, if we adopt the RGB mass (0.82 $M_{\odot}$) for all targets, 
as it is often done because of the difficulty to observationally distinguish 
between RGB and AGB stars, the values of [Fe II/H] increase by only 0.03 dex 
with respect to estimate obtained assuming 0.68 $M_{\odot}$.
Hence, the precise value of the adopted mass (within a reasonable mass range) cannot reconcile the difference 
between [Fe I /H] and [Fe II/H].

As additional check, for each AGB stars, the stellar mass has been varied until 
the ionization equilibrium was satisfied. 
The derived values range from $\sim$0.2 and $\sim$0.5 $M_{\odot}$: such masses are too low with 
respect to the mass distribution of the HB stars derived by \citet{gratton10}. 
In particular, for the stars \#89, \#181 and \#240, that exhibit the largest difference between 
[Fe I/H] and [Fe II/H] ($\sim$--0.2 dex), a satisfying ionization equilibrium can be reached only
with masses smaller than 0.2-0.25 $M_{\odot}$, which are very unlikely values 
for globular cluster AGB stars.

\section{Conclusions}

We demonstrated that the observed intrinsic star-to-star Fe scatter 
in the GC NGC~3201 is due to unaccounted NLTE effects in the spectroscopic analysis of some AGB stars included 
in the sample. These stars suffer from NLTE effects driven by the iron overionization, 
a mechanism that affects mainly the less abundant species like Fe~I, but has no a
significant effect on the dominant species (e.g. Fe~II).
When the gravity of these stars is obtained spectroscopically, forcing to have 
the same abundance from Fe~I and Fe~II lines, the derived [Fe/H] abundance 
turns out to be under-estimated.

Our findings confirm the conclusion by \citet{lp14} that the chemical analysis of samples of stars 
including both AGB and RGB stars, and based on spectroscopic gravities, can lead to 
spurious broadening of the iron distribution.

{\sl We conclude that NGC~3201 is a normal GC, without evidence of intrinsic iron scatter.}
In light of this result, it is not necessary to suppose that NGC~3201 was more massive in the past 
to retain the SN ejecta, as invoked by \citet{simmerer13}.


\acknowledgements  
We warmly thank the anonymous referee for suggestions that helped improving the paper.
This research is part of the project COSMIC-LAB (http://www.cosmic-lab.eu) funded 
by the European Research Council (under contract ERC-2010-AdG-267675).


\end{document}